# A probable Keplerian disk feeding an optically revealed massive young star

Anna F. McLeod[1,2], Pamela D. Klaassen[3], Megan Reiter[4], Jonathan Henshaw[5,6], Rolf Kuiper[7], Adam Ginsburg[8]

[1]Centre for Extragalactic Astronomy, Department of Physics, Durham University, South Road, Durham DH1 3LE, UK
[2]Institute for Computational Cosmology, Department of Physics, University of Durham, South Road, Durham DH1 3LE, UK
[3]UK Astronomy Technology Centre, Royal Observatory Edinburgh, Blackford Hill, Edinburgh, EH9 3HJ, UK
[4]Department of Physics and Astronomy, Rice University, 6100 Main St - MS 108, Houston, TX 77005, USA
[5]Astrophysics Research Institute, Liverpool John Moores University, IC2, Liverpool Science Park, 146 Brownlow Hill, Liverpool L3 5RF, UK
[6]Max Planck Institute for Astronomy, Königstuhl 17, D-69117 Heidelberg, Germany
[7]Faculty of Physics, University of Duisburg-Essen, Lotharstraße 1, D-47057 Duisburg, Germany
[8]Department of Astronomy, University of Florida, PO Box 112055, Gainesville, FL, 32611-2055, USA

**The canonical picture of star formation involves disk-mediated accretion, with Keplerian accretion disks and associated bipolar jets primarily observed in nearby, low-mass young stellar objects (YSOs). Recently, rotating gaseous structures and Keplerian disks have been detected around a number of massive (M > 8 $M_\odot$) YSOs (MYSOs)[1,2,3,4], including several disk-jet systems[5,6,7]. All of the known MYSO systems are located in the Milky Way, and all are embedded in their natal material. Here we report the detection of a rotating gaseous structure around an extragalactic MYSO in the Large Magellanic Cloud. The gas motions show radial flow of material falling from larger scales onto a central disk-like structure, the latter exhibiting signs of Keplerian rotation, i.e., a rotating toroid feeding an accretion disk and thus the growth of the central star. The system is in almost all aspects comparable to Milky Way high-mass young stellar objects accreting gas via a Keplerian disk. The key difference between this source and its Galactic counterparts is that it is optically revealed, rather than being deeply embedded in its natal material as is expected of such a young massive star. We suggest that this is the consequence of the star having formed in a low-metallicity and low-dust content environment, thus providing important constraints for models of the formation and evolution of massive stars and their circumstellar disks.**

The lack of observations of optically revealed MYSOs is a consequence of the rapid timescales on which massive stars form: they form in heavily embedded regions full of gas and dust, such that the accretion phase typically occurs prior to the star having time to become exposed due to stellar feedback, whether internal or external. The primary reason for the lack of observations of extragalactic accretion disks around forming stars has been the limited spatial resolution of both ground- and space-based observatories.

At a distance of 50 kpc, the LMC has proven to be a convenient environment to search for the extragalactic counterparts of the known Milky Way accreting MYSOs, with the recent detection of sub-parsec scale molecular outflows[8,9], as well as the discovery of HH 1177, a collimated bipolar jet driven by an MYSO[10]. While the detection of a molecular outflow from a forming star does not necessarily imply the presence of an accretion disk, collimated jets are generally taken as clear signposts for ongoing disk accretion. To date, there has been no direct detection of a rotating circumstellar Keplerian (accretion) disk or toroid in an external galaxy, making the HH 1177 star/jet system an ideal target to search for these. The Atacama Large Millimeter Array (ALMA) now enables the high sensitivity and high angular resolution observations needed to detect and resolve rotating circumstellar gas in extragalactic MYSOs.

The rotating structure reported here is feeding the central star of the HH 1177 system, which was previously detected in optical integral field spectroscopic observations[10]. The system is located in the LMC star-forming region N180, a classical HII region photoionized by the OB association LH 117[11,12]. HH 1177 comprises a bipolar (externally) ionized jet with a total (projected) extent of 11 pc (Figure 1), originating from a central source classified as an MYSO with a mass of 12 M$_\odot$ estimated from infrared SED fitting[13], indicating that the central star is likely a B-type rather than an O-type star at its current evolutionary stage. The central star of HH 1177 has formed at the tip of a pillar-like molecular cloud structure, oriented pointing toward (in projection) three massive stars of the LH117 star cluster. HH 1177 remains the only known extragalactic MYSO/jet system, providing a unique laboratory for MYSO formation and evolution studies in an external galaxy. In 2019 and 2021 the central star of HH 1177 was targeted with ALMA Band 7 (275 – 373 GHz) in two different configurations, capturing emission on different size scales. Combined, these observations result in a continuum angular resolution of 50 mas × 40 mas (2500 au at the distance of the LMC). The ALMA observations covered the molecular lines $^{12}$CO (J = 3-2), $^{13}$CO (J = 3-2), CS (J = 7-6), and CH$_3$CN (J = 18-17). In addition to the molecular lines, a fourth spectral window centered on 0.870 mm was devoted to continuum observations, which show a compact, marginally resolved, continuum source at the location of the central star of HH 1177. The continuum source (shown as contours in Figure 2, top panels) has a peak flux of 0.28 mJy beam$^{-1}$, an integrated flux of 0.34 mJy, and a deconvolved size of 50 mas × 32 mas at a position angle of 60°.

While we detect no CH$_3$CN emission (and there are no spurious line detections of other species within the spectral windows), the velocity maps of CS and $^{13}$CO (Figure 2; the velocity maps were obtained via a multi-component spectral decomposition, necessary due to the presence of multiple velocity components which complicate the analyses, see Methods), tracing the dense gas kinematics, show a distinct velocity gradient almost perpendicular to the red and blue lobes of the ionized jet (the jet orientation is indicated by the solid black line in Figure 2). The red and blue jet lobes have P.A. of 144° and -32° with respect to the central source, respectively. We use a P.A. of 55°, i.e., ≈ 90° with respect to the jet, to extract position-velocity (PV) diagrams for both lines.

The PV diagrams (Figure 3, bottom panels) show the characteristic "butterfly" shape typical of a rotating structure, with higher gas velocities closer to the centre and consistent with velocities $v_{rot} \propto R^{-\alpha}$, with $\alpha > 0$. For both CS and $^{13}$CO, the kinematics along the outer edge of the emission (dotted dark blue line in Figure 2; see Methods) exhibits Keplerian motion ($\alpha = 0.5$) in the innermost regions (offsets < 0.12 arcsec, i.e., 0.029 pc or approximately 6000 AU), and with pure free fall ($\alpha = 1$) in the outer parts of the structure. Further insight into the gas kinematics is gained from the channel maps (Extended Data Figures 1 and 2): with systemic velocities of 225.3 km/s and 225.2 km/s derived for CS and $^{13}$CO, respectively, the channel maps confirm that the location of the highest velocity gas is near the central star. The kinematics are therefore indicative of the molecular lines tracing a rotating gaseous structure around the MYSO launching the jet of HH 1177.

We further explore the kinematics of the rotating gas by fitting the velocity profiles of the outer envelope with power-laws[14,5] (see Methods and Extended Data Figure 3). For both tracers, we can exclude radial infall ($v \propto R^{-1}$), while a Keplerian profile is consistent with the data, yielding a central star (enclosed) mass, $M_*$, for the central source of 14.7±0.8 $M_\odot$ and 19.5±1.3 $M_\odot$ (respectively for CS and $^{13}$CO). The derived enclosed stellar mass is broadly consistent with $M_* = 12 M_\odot$ derived via SED fitting[13].

While consistent with Keplerian rotation over the extent of the rotating structure, the best-fit power-law with varying index (as opposed to a fixed index) yields an index $\alpha$ which is considerably smaller than $\alpha = 0.5$, indicating a combination of different kinematics (see Extended Data Figure 3). With Keplerian rotation curves describing the kinematics at smaller separations better than those at larger separations, we distinguish between an inner and an outer part of the rotating structure, using the offset of 0.12 arcsec (6000 AU, see Methods) as a critical radius. In the inner region, i.e., $R < 6000$ AU, the kinematics are best described by Keplerian rotation (Figure 3). This is particularly evident in CS given the better sampling of the data in this line. However, the data is insufficient to distinguish between the adopted central star mass of 15 $M_\odot$ (obtained from the emission over the entire extent of the rotating structure) and a best-fit Keplerian mass of ~ 9.6 $M_\odot$ (obtained from only the emission within 6000 AU). Large Keplerian disks with radii up to 1000-3000 AU have been detected around Milky Way MYSOs[2,3,6], and all except for one of the most likely Keplerian disk candidates around B-type protostars (i.e., Galactic counterparts to the HH 1177 central star), also have spatially resolved disk radii in the ~1000-3600 AU regime[15]. Hence, while a factor of 2-6 larger than the aforementioned disk radii, 6000 AU is of the same order of magnitude, and in what follows, we use the outer radius of 6000 AU of what we refer to as the "inner region" of the rotating structure as a "disk radius", $R_d$. However, we caution that the Keplerian disk is not well resolved, and the continuum source (likely tracing the true disk in the inner regions which material from the larger scales is falling onto) is only marginally resolved; we adopt $R_d = 6000$ AU for further analysis as a useful limiting case.

The presence of a highly collimated bipolar jet and of rotating material transitioning to motions consistent with Keplerian dynamics around the central star of the HH 1177 system support the picture of this MYSO forming via disk-mediated accretion and is in line with recent numerical models of disk-jet system around massive protostars[16,17]. With all other known accreting MYSOs being in the Milky Way, the detection of the HH 1177 disk offers the opportunity to empirically analyse how the formation of massive stars might differ in an environment with comparatively lower metal and dust contents. To compare the HH 1177 system to Galactic counterparts, we compute the disk mass and accretion rate, and analyse the disk stability.

We find a disk gas mass, $M_g$, derived from the continuum image (assuming a gas-to-dust ratio of 380 for the LMC[18] to convert between a dust mass and gas mass) to be between ~1.8 $M_\odot$ and ~3.9 $M_\odot$ (assuming a temperature of 100 K and 50 K, respectively, see Methods section), which corresponds to ~ 12-26% of the stellar mass, the former being consistent, within errors, with the value of ~0.5 $M_\odot$ obtained from SED fitting of near-infrared observations of this source[13]. In the relation between the gas mass of circumstellar structures (whether disks or toroids) and the enclosed stellar mass[15], about half of the known high- (and intermediate-) mass protostars are situated above the 1-to-1 line. In this parameter space, the handful of known B-type stars with the highest likelihood of hosting Keplerian disks tend to fall below the line, having $M_g < 0.3 M_*$ (Figure 4). HH 1177 system is in the same region of parameter space. This further supports the picture of the central star being a B-type star with a Keplerian circumstellar disk. When taking into account the disk radius (with the caveat that we are likely overestimating it), the ratio $M_g / M_*$ is comparable to what is found for Galactic counterparts with radii $R > 2000$ AU (Figure 4). This is further supported by the disk-averaged surface density ($\Sigma = M_g / \pi R_d^2$) of HH 1177 and the Galactic disks with $R > 2000$ AU being of the same order of magnitude, as the disk mass increases with $R_d$.

With the disk having a mass $M_g < 0.3\,M_*$, the expectation of the disk being gravitationally stable[19] is supported by a Toomre parameter $Q > 1$ (at $R_d = 6000$ AU). While this is consistent, within a factor ~3, with Galactic Keplerian disks around B-type stars (Figure 5), numerical work shows that disk fragmentation is dependent on metallicity, with strongly unstable disks[20] resulting in higher fragmentation at lower metallicity[21]. Additionally, circumstellar disks tend to begin their lives with an unstable epoch, followed by an epoch of stability later on[22], and $Q$ is a local parameter that generally decreases with disk radius. We would therefore expect the HH 1177 disk to have a lower Toomre $Q$ parameter compared to the MW B-type stars with Keplerian disks, especially at a radius of 6000 AU. However, the opposite is the case. While the overestimation of the disk radius and the large uncertainties stemming from the temperature assumptions play a role in producing a large value for $Q$, a possible explanation for this super-stable disk is that the HH 1177 system is exposed to the stronger radiation field at lower metallicities of the driving source compared to a MW star, such that the higher photon flux might contribute to maintaining the high disk temperature and thus preventing fragmentation[23]. While there are about 14 O-type stars in N180[12], their contribution towards supporting the disk via external heating is negligible (see Methods).

The growth of the central star is fueled by accretion, and for low-mass pre-main sequence (PMS) stars the mass accretion rate is proportional to the stellar mass as $\dot{M}_{acc} \propto M_*^{1.5\text{-}2.0}$ [24]. Beltrán & de Wit[15] compared the mass accretion rates of embedded and revealed stars across a mass range spanning from low-mass PMSs to embedded high-mass protostars, therefore also sampling different evolutionary stages. They find a positive trend broadly consistent with $\dot{M}_{acc} \propto M_*^2$, but that embedded YSOs have systematically higher mass accretion rates compared to PMSs, suggesting a decrease of $\dot{M}_{acc}$ with evolutionary stage. However, this remained inconclusive due to the lack of optically revealed high-mass (M > 10 M$_\odot$) stars with accretion disks. The previously derived mass accretion rate of ~9.5 × 10$^{-6}$ M$_\odot$ yr$^{-1}$ (derived assuming that $\dot{M}_{acc} = 3.3\,\dot{M}_{jet}$ [10]) places the HH 1177 system in that previously existing gap in the parameter space (Figure 5). It reconciles the apparent shift to higher $\dot{M}_{acc}$ of embedded high-mass sources compared to optically revealed Herbig Ae/Be stars, therefore strengthening the likely implication that the same processes are driving accretion for all central stars and that the formation mechanism of high-mass stars is a scaled-up version of low-mass star formation.

In summary, the likely Keplerian disk of the central high-mass star driving HH 1177 is, in almost all aspects, similar to its MW counterparts. However, it stands out for two reasons. The first is that rather than being embedded in its natal molecular cloud like all other known MYSOs, it is the only optically revealed high-mass YSO. The second reason is the stability of the disk. We suggest that both of these are due to the low-metallicity and low-dust content of the birth environment of the HH 1177 system, which impact the physical processes governing the optical depth of the surrounding matter. At lower metallicities, stars yield a higher number of photons in the EUV part of the spectrum, and due to less efficient cooling, the temperature of the surrounding gas is higher. This results in stronger thermal pressure feedback as well as higher photo-evaporation rates, and the lower dust content decreases the overall continuum optical depth. Conversely, while we cannot exclude that the outermost radii of the disk are unstable, internal irradiation is likely responsible for maintaining a high disk temperature, stabilizing it against fragmentation. Determining the consequences of these environmental differences provides important constraints for our theoretical understanding of the formation and evolution of massive stars and their circumstellar disks.

**Acknowledgements**

J. H. gratefully acknowledges financial support from the Royal Society (University Research Fellowship; URF/R1/221620). R. K. acknowledges financial support via the Heisenberg Research Grant funded by the German Research Foundation (DFG) under grant no. KU 2849/9. AG acknowledges support from the NSF under AST 2008101, 2206511, and CAREER 2142300.

**Author contributions**

A. F. M. is the principal investigator of the ALMA observations used for this work and is the main contributor of the manuscript. P. K. performed the data reduction and contributed to the manuscript text. A. G. provided the scripts used to perform the outer envelope analysis. J. H. performed the SCOUSE and ACORNS analysis. R. K. and M. R. provided the input concerning the theoretical and observational aspects (respectively) of the interpretation. All coauthors commented on the manuscript.

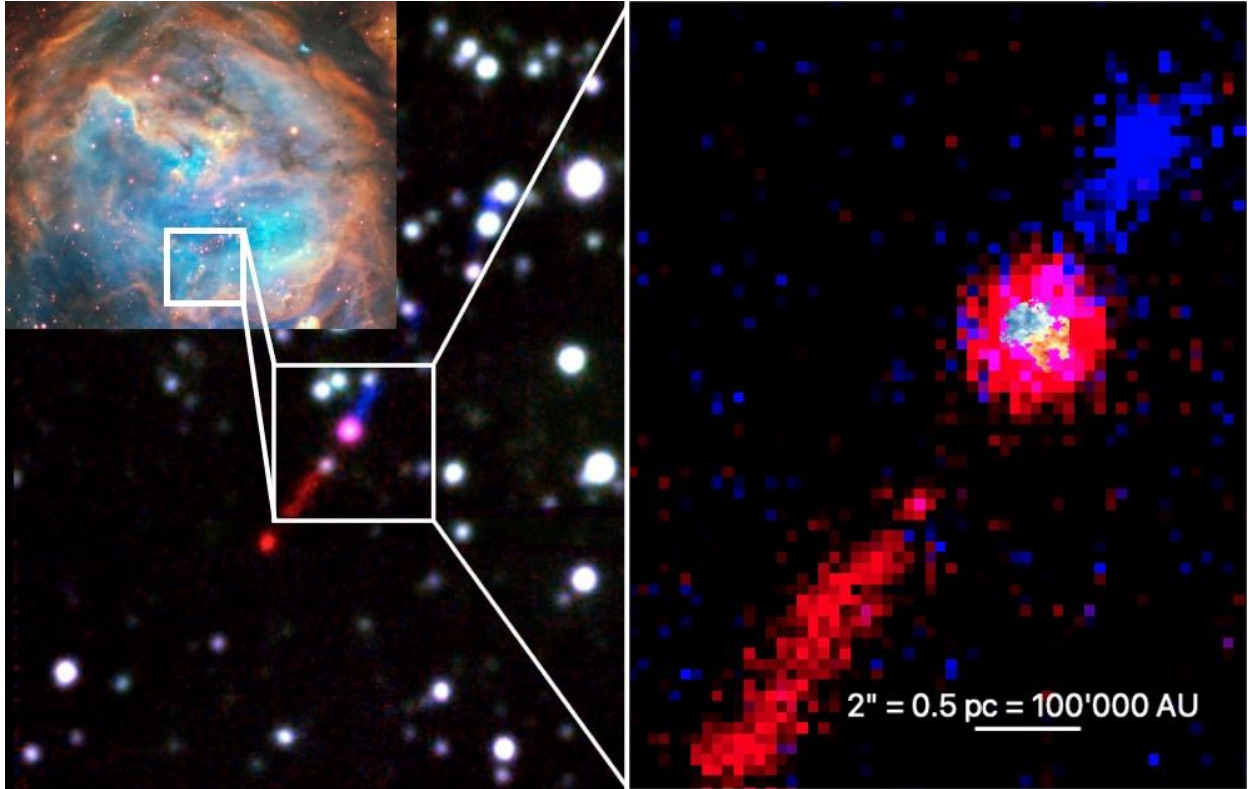

**Figure 1. RGB composites of the star-forming region N180 and the jet.** *Left panel, inset: three-color composite of the star-forming region N180 in the Large Magellanic Cloud (red = [SII]6717, green = Hα, blue = [OIII]5007), observed with the MUSE instrument on the Very Large Telescope[10]. Main left panel: the red- and blue-shifted wings of the Hα emission line highlighting the externally irradiated HH 1177 jet emerging from the central star. Right panel: same as the main left panel but continuum-subtracted and zoomed in onto the driving star of the HH 1177 system, with the CS velocity map overlaid showing the rotating molecular gas at the location of the central star (note that, for illustrative purposes, the CS velocity map has been enlarged by a factor ~4).*

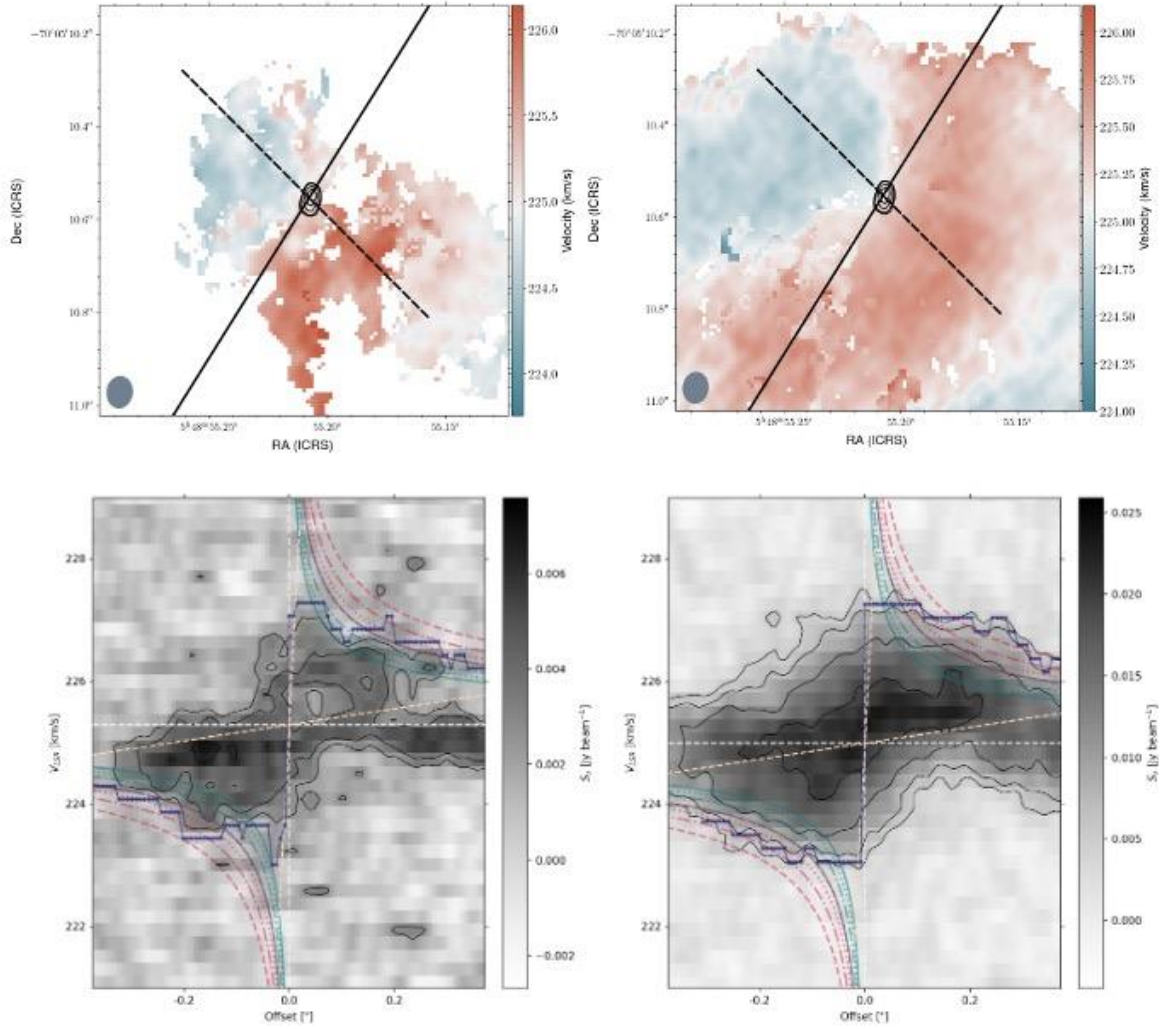

**Figure 2. Kinematics of the molecular gas.** *Velocity maps of CS and $^{13}CO$ (left and right top panels, respectively) derived from SCOUSEPY and ACORNS decompositions (see Methods). The shown velocity corresponds to the local standard of rest velocity, $v_{LSR}$. Black contours indicate the location of the continuum source and are shown at 0.17, 0.18, 0.21, 0.25, and 0.28 mJy/beam. The black dashed line corresponds to the 0".75 and 55 °P.A. slit used to extract the PV diagram, i.e., approximately perpendicular to the optical jet of the HH 1177 system, indicated by the solid black line[10]. Position-velocity (PV) diagrams extracted along the slits shown in Figure 2 for CS and $^{13}CO$ (left and right bottom panels, respectively). The dashed white lines correspond to the adopted central velocity and position of the source ($v_{LSR} = 225.3$ km/s, 05:48:55.2099 -70:05:10.579). The dashed salmon line indicates the rotating structure's assumed inner and outer limits, 1000 and 25000 AU, respectively. The dark blue points and line indicate the outer envelope (i.e., the outer edge[14,5]). The sequence of teal and pink lines show, respectively, the Keplerian (v ∝ $\sqrt{M/R}$) and free fall (v ∝ $\sqrt{2M/R}$) velocity curves for a central source of 10, 12, 15, and 20 $M_\odot$ (solid, dotted, dot-dashed, dashed, respectively). The series of displayed central star masses encompasses a slightly less massive option to the value of 12 $M_\odot$ inferred from Spitzer data[13], as well as masses of the order of what is derived in this work (see Methods).*

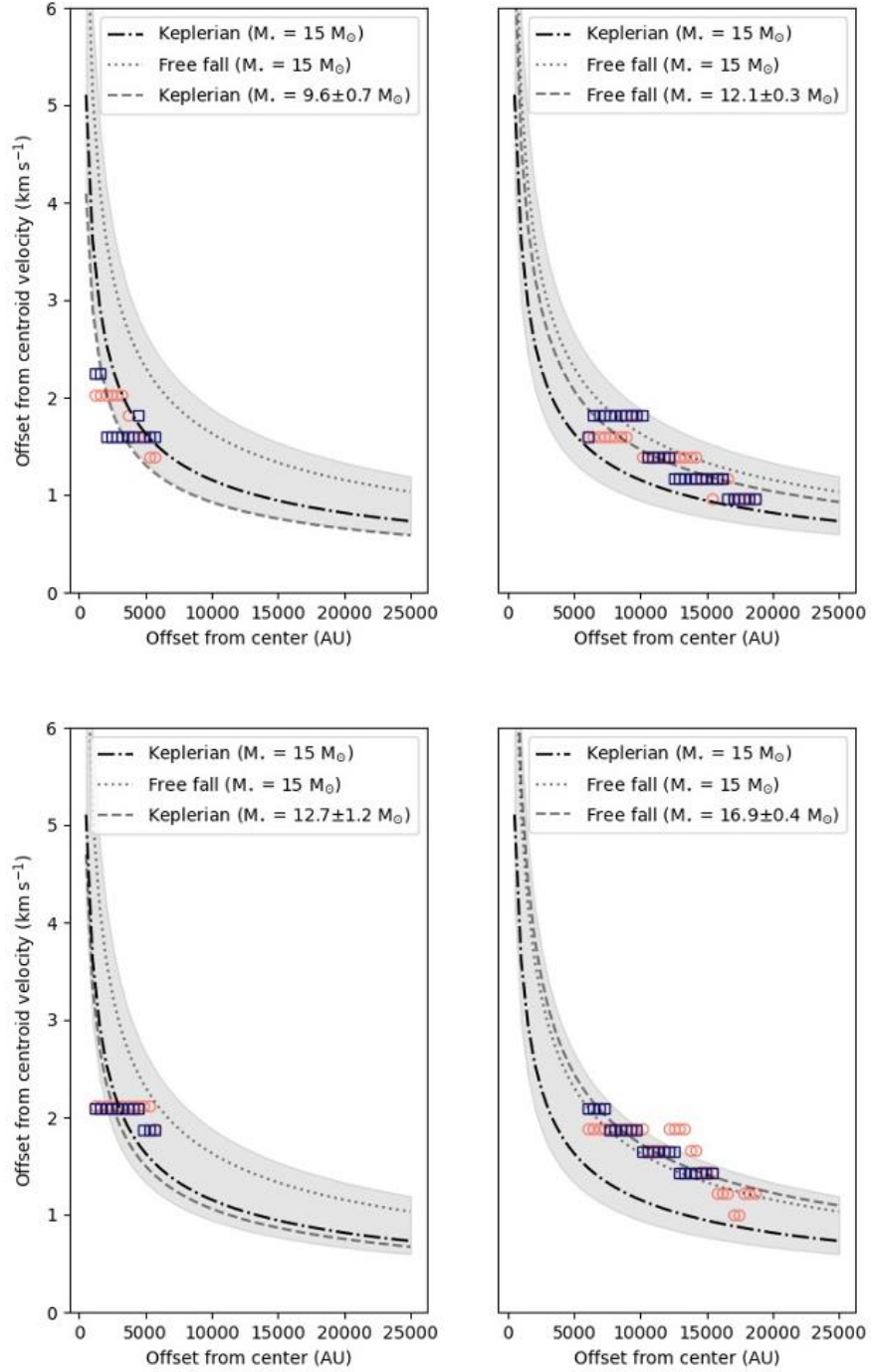

**Figure 3. Kinematics of the inner and outer parts of the rotating structure.** *Kinematics in the inner (R < 6000 AU, left panels) and outer (R > 6000 AU, right panels) regions of the rotating gas. Shown are the redshifted (orange circles) and blueshifted (dark blue squares) radial velocity profiles of the outer envelope from Figure 2 (bottom panels) for CS and $^{13}CO$ (upper and lower panels, respectively). Dot-dashed and dotted curved correspond to Keplerian and free fall kinematics for an assumed fixed central source mass of 15 $M_\odot$, while the dashed curve corresponds to the best-fit Keplerian (inner regions) or free fall (outer regions) curves, showing that the inner regions are best described by Keplerian-like motions, while the outer regions are best described by free fall. Masses of the central source derived from the Keplerian and free fall fits are indicated in the legend. The shaded area indicates Keplerian rotation for the mass range 10 $M_\odot$ < M < 40 $M_\odot$.*

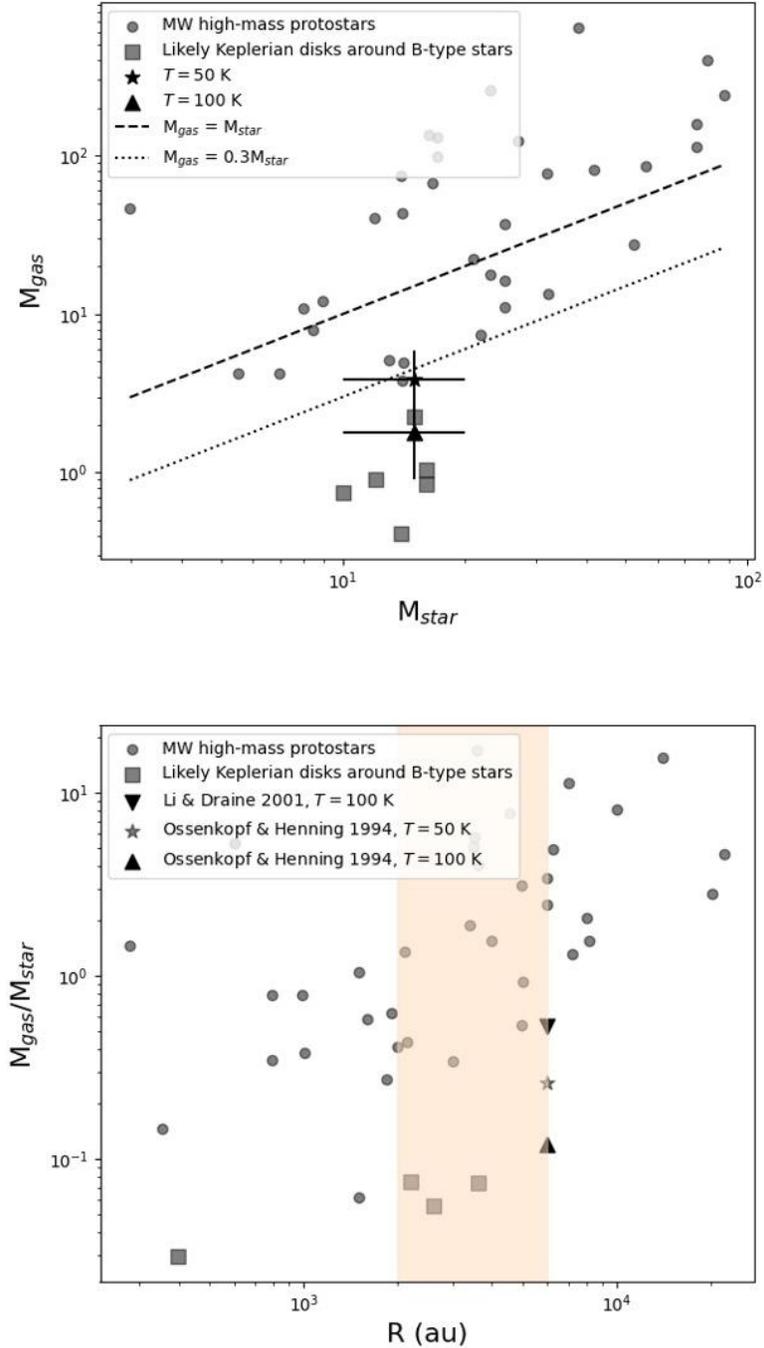

***Figure 4. Comparison with Milky Way objects.*** *Comparison of the HH 1177 disk to Galactic high-mass protostars (circles), with data from Beltrán & de Wit[15]. In both panels, squares correspond to the most likely Keplerian disks around (Galactic) B-type stars (all of which but one have disk radii >1000 AU) assuming Ossenkopf & Henning[25] opacities. The upper panel shows $M_{gas}$ obtained by assuming $\kappa_\nu$ from Ossenkopf & Henning and a disk temperature of 100 K (triangle) and 50 K (star) together with associated errors, see Methods. The lower panel illustrates different resulting $M_{gas}/M_{star}$ values for the HH 1177 source when assuming $\kappa_\nu$ from Ossenkopf & Henning or from Li & Draine[26] (see Methods). The star and triangle are as per the upper panel. The shaded area spans disk radii of 2000 AU up to 6000 AU, i.e., from radii of similar structures to the (upper limit) radius of the HH 1177 source. Not all objects indicated with squares in the upper panel are in the lower panel, as some of these do not have known stellar masses.*

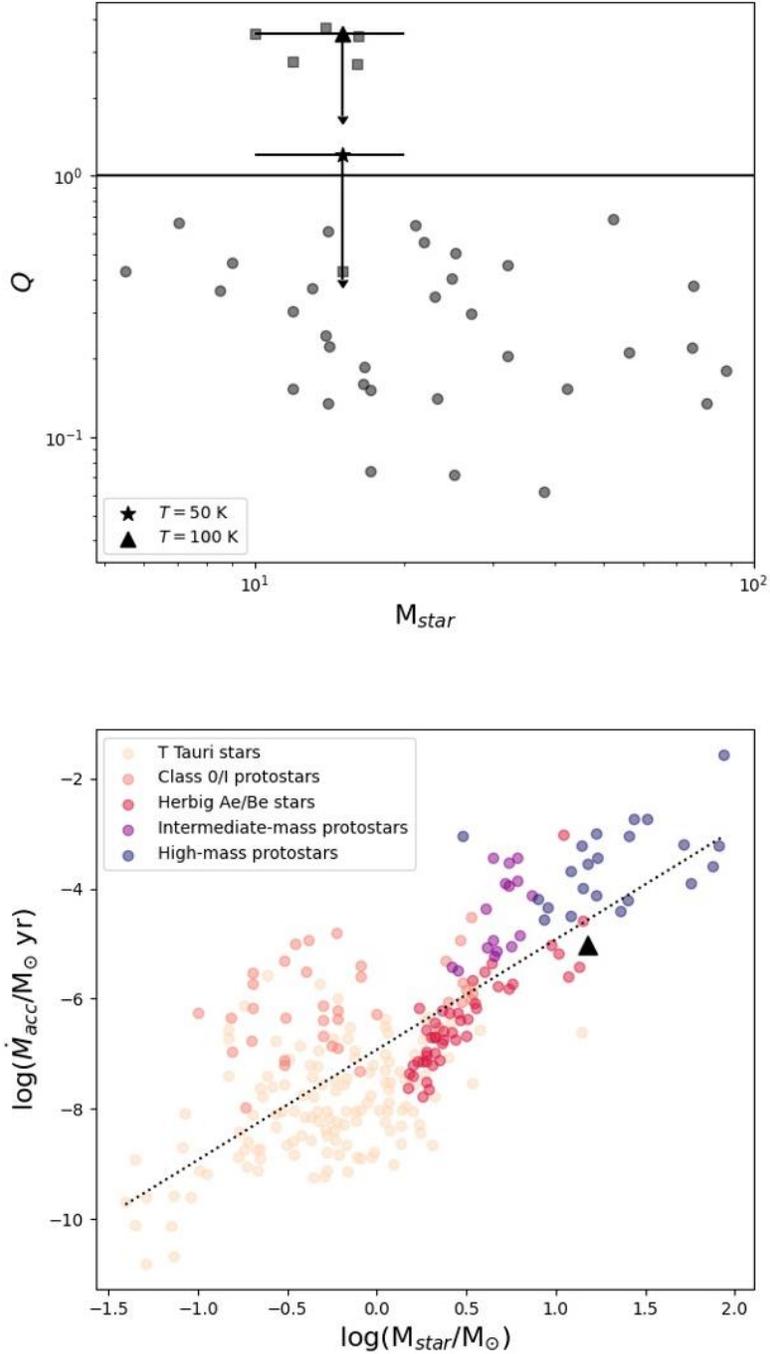

**Figure 5. Comparison with Milky Way objects (continued).** *Comparison of the HH 1177 disk to Galactic high-mass protostars (circles), with data from Beltrán & de Wit[15]. The upper panel shows the Toomre Q parameter of the HH 1177 disk as a strict upper limit, with squares corresponding to the most likely Keplerian disks around (Galactic) B-type stars assuming Ossenkopf & Henning[25] opacities and disk temperature of 100 K (triangle) and 50 K (star) together with their associated errors, see Methods. The lower panel shows the relation between mass accretion rate and stellar mass for a variety of young stellar objects. The dotted line corresponds to a $\log{(\dot{M}_{acc})} \propto 2\times \log(M_{star})$ relation. The HH 1177 system is consistent with being an optically revealed, young high-mass star accreting from a stable circumstellar disk.*

## METHODS

### Data Reduction

ALMA observations for this project were taken as part of project 2019.1.00756.S. To resolve the expected Keplerian rotation, we required a spatial resolution of at least 0.05" within a pillar that emits on 4" scales. To capture emission on those large size scales with that resolution required two ALMA configurations. The larger scale observations were taken on 7 and 8 October 2019, and the high-resolution data were observed on 25 and 21 October and 1 November 2021. The two sets of observations were reduced using the standard ALMA pipeline and combined in the *uv* plane using the CASA command *concatvis*. Both data sets were imaged and cleaned independently to verify their signal and noise levels. The combined data set was successfully imaged, but the cleaning algorithms failed; increasing the noise and failing to find the signal, despite trying multiple cleaning methods such as multi-scale cleaning, using various clean boxes, and manual cleaning. Thus, for this paper, we use the dirty images and note that the flux levels derived from the combined dirty images are consistent with those of the low-resolution clean images and higher than those of the high-resolution clean images, which suffer from spatial filtering. Morphologies, kinematics, and mass estimated from the kinematics are not affected.

### SCOUSEPY and ACORNS Decomposition

Due to the presence of multiple velocity components (present in both $^{13}CO$ and CS), which vary between one to three components on a pixel-by-pixel basis, we use the multi-component spectral line decomposition algorithm SCOUSEPY[27,28] (Semi-automated multi-COmponent Universal Spectral-line fitting Engine) to fit the spectral line data. In brief, SCOUSEPY uses a semi-automated step-by-step approach towards producing a parametric, pixel-wise, multi-component description of spectroscopic data cubes. We performed a Gaussian decomposition of the $^{13}CO$ and CS data cubes. To define the coverage for the SCOUSEPY decomposition we masked the data cubes at a level of 0.007 Jy/beam and 0.0035 Jy/beam, respectively. We generated spectral averaging areas (SAAs) of size 20 and 10 pixels, resulting in 356 and 207 SAAs, respectively, with the larger size of the $^{13}CO$ SAAs reflecting the more extended emission. Of these SAAs, a total of 25 and 41 were decomposed manually, while the remaining SAAs made use of SCOUSEPY's derivative spectroscopy methodology (Henshaw et al. in prep.). Of the 34660 and 5600 spectra contained within the $^{13}CO$ and CS data cube coverage, 34560 and 5314 have model solutions. Of these model solutions, the fractional number of pixels requiring multi-component models is small (of the order 10%), however, the maximum number of components identified within a single spectrum can be as high as 4, justifying the need for Gaussian decomposition.

Given the presence of multiple velocity components in our data, we next used ACORNS (Agglomerative Clustering for ORganising Nested Structure) to cluster the extracted components into velocity coherent features. ACORNS is based on a technique known as hierarchical agglomerative clustering, which generates a hierarchical system of clusters of data connected in n-dimensional space via selected properties. In the case of our parametric description of the velocity components output from SCOUSEPY, we clustered these data based on the separation between data points in physical space, their velocity centroids, and their velocity dispersion. We set the linking lengths for clustering based on the observational limitations of our data. Specifically, we define the minimum cluster size as being 25 pixels and set the linking length in velocity and velocity dispersion to be twice the spectral resolution of our data. From the resulting hierarchy, we extracted the largest clusters identified by ACORNS, which are displayed in Figure 2 (top panels). While most of the pixels display a single component (> 92% in $^{13}$CO and > 88%), all pixels are shown in Figure 2 (top panels).

## Outer envelope analysis, mass of central source, and gas kinematics

The mass estimate of the star is performed according to the method described in Seifried et al. (2016)[14], under the assumption that the motions of the disk follow Keplerian rotation. Briefly, this method consists in first estimating the mean level of noise, $\sigma_{rms}$, in the outer parts of the PV diagrams (i.e., where no emission is present) and, based on this, for each radial offset position in the PV diagram starting from the offset position at highest velocity (or lowest, for the opposite quadrant), identify the first pixel with emission above the adopted threshold value. The velocity of this pixel corresponds to the maximum (minimum) rotation velocity at that offset radius. This results in the identification of the outer edge of the rotating structure in the form of the maximum rotation velocity as a function of positional offset. This can then be used to constrain the mass of the central source based on the best fit (Keplerian) curve to the extracted data points.

We find mean levels of noise $\sigma_{rms}$ of 1.5 mJy and 0.4 mJy for $^{13}$CO and CS, respectively. Here, we adopt a threshold of $5\sigma_{rms}$, which produces the "cleanest" velocity vs. offset data (middle panels in Extended Data Figure 4). Figure 2 (bottom panels) shows the PV diagrams for the two species and the resulting outer edge of the structure, together with curves expected for Keplerian (magenta lines) and free fall (yellow lines) motions around a central source of different masses.

The red- and blueshifted sides of the outer envelope are folded into a single plot in Extended Data Figure 3 to further analyse the kinematics of the structure. We perform power law fits the PV diagram of the form $v = \beta R^{-\alpha}$, where $\alpha = 0.5$ corresponds to Keplerian and $\alpha = 1$ to infall motions. Additionally, we perform a fit in which the exponent $\alpha$ is allowed to vary. While the data is clearly not described by infall, the fits demonstrate that the kinematics are likely a combination between Keplerian (or sub-Keplerian, with $\alpha \sim 0.2$) rotation and free fall. The best-fit central source mass for Keplerian rotation is (19.5±1.3) and (14.7±0.8) M$_\odot$ for $^{13}$CO and CS, respectively. In the following analyses, we assume a central star mass of 15 M$_\odot$ for simplicity. We note, however, that the derived stellar mass is a lower limit, due to the effect of the inclination angle $i$ on the mass estimate, such that lower masses are derived via the outer envelope calculation in systems that are increasingly deviating from an edge-on viewing angle[14]. With an inclination of about 73° [13], the HH 1177 system is close to edge-on, we expect the derived mass to be smaller by a factor of about 0.9 (given that the fitted mass scales as $cos^2(90°-i)$).

Further, as can be seen in Figures 2 and Extended Data Figure 3, the resulting outer envelope is more consistent with Keplerian motion at smaller offsets ($\lesssim 0.15$ arcsec, i.e., in the vicinity of the continuum source), and with free fall motion at larger offsets. To further illustrate this, we separate the outer envelope data points into two samples describing the regions closer ($R < R_{crit}$) and further away ($R > R_{crit}$) from the center. We adopt a critical radius, $R_{crit}$, of 6000 AU, i.e., twice the mean angular resolution, and perform two fits: first, we assume a central source mass of 15 M$_\odot$ and fit a Keplerian and a free fall model to the two regimes; second, we let the central source mass vary and fit a Keplerian model to the inner region, and a free fall model to the outer region. This is summarised in Figure 3. For CS, while we cannot rule out free fall motions with a central mass of $\lesssim 6$ M$_\odot$, the inner regions are clearly better described by Keplerian rotation with a central star in roughly the mass regime expected from the literature. The opposite is true for the outer regions traced by CS: while we cannot rule out Keplerian rotation with a central mass $\gtrsim 24$ M$_\odot$, free fall with a central source of the expected mass is a good description. While less predictive, the same argument is valid for $^{13}$CO. We, therefore, suggest that the gas in the detected rotating structure is free-falling from the outer regions onto a central disk where the kinematics are (sub-)Keplerian. The scale of a few thousand AU at which the transition to a Keplerian disk occurs are consistent with the size of Keplerian accretion disks observed around massive stars[15].

**Mass of inner disk and outer envelope**
The area used for to derive a source flux for the mass calculation below was derived using the CASA task *imfit*. This task fits a 2D Gaussian to the emission in the continuum image and reports the area (deconvolved from the synthesised beam) and enclosed flux.

$$M_{dust} = \frac{d^2 F_\nu}{\kappa_\nu B_\nu(T_{dust})}$$

where $d$ is the distance to the source, $F_\nu$ is the integrated flux of 0.34 mJy of the continuum source (as described above), $\kappa_\nu$ the dust opacity per unit mass at a frequency $\nu$, and $B_\nu$ the Planck function at a temperature $T_{dust}$.

The main sources of uncertainty in the estimation of the dust mass are stemming from $\kappa_\nu$, and $T_{dust}$[15]. Here, we use a dust opacity from Ossenkopf & Henning[25] (see below), which differs from Johnston et al.[2] who use opacities from Draine[29] (which are systematically lower than the Ossenkopf & Henning values and likely more suitable for diffuse clouds[6]), but in line with other studies of disks around massive stars[30,3,6], and consistent with the Beltrán & de Wit[15] review for comparative purposes. All opacities assume Galactic metallicity.

We compute the disk mass and subsequent parameters assuming two temperatures. First, we assume a temperature of 100 K, based on the properties of other likely Keplerian disks of similar central stars and sizes[31,32]. Second, we derive a disk temperature from radiative equilibrium (equating cooling and heating rates[33]),

$$T_{dust} = \sigma^{-\frac{1}{4}}(1-\alpha)^{1/4}\left(\frac{L_{star}}{4\pi R^2}\right)^{1/4}$$

where $\alpha$ is the albedo (assumed to be 0.6), and $L_{star}$ and $R$ are the luminosity of and the distance from the central star, respectively. Based on the central source parameters (temperature, radius) derived via SED fitting[13], we assume $L_{star} \sim 1.9 \times 10^4$ L$_\odot$ based on the best comparable LMC stellar atmosphere model (model 28-40 in Hainich et al.[34]), luminosity which is comparable to that quoted for Galactic B-type stars bearing Keplerian disks[15]. Inserting $R = 6000$ AU yields $T_{dust} \approx 50$ K. While there are about four O-type stars within a (projected) radius of 12 pc[12], their contribution to heating the disk is negligible (their contribution to the equation above would be in the form of an added $L_{star}/4\pi D^2$ term, with $D$ being the projected distance to the external sources). We note that the disk temperature is radius-dependent, such that assuming a single temperature leads to large uncertainties in both the disk mass and the stability calculation described below.

We adopt a distance of 49.59 kpc to the LMC[35] and a dust opacity at 0.87 mm of $\sim 2.5$ cm$^2$ g$^{-1}$ (assuming grains with thin ice mantles and coagulation at $10^8$ cm$^{-3}$). The dust mass is then converted into a gas mass as $M_{gas} = \text{GDR} \times M_{dust}$, where GDR is the gas-to-dust ratio which for the LMC is GDR = $380^{+250}_{-130}$[18]. To compute the uncertainty on the derived gas mass, we assume a 20% error on the flux measurement and propagate these together with the GDR uncertainties. The derived disk gas mass is thus 1.8±0.9 M$_\odot$ at 100 K, and 3.9±2.0 M$_\odot$ at 50 K. These corresponds to

≈ 12% and 26% of the stellar mass (assuming $M_{star} \sim 15$ M$_\odot$, see above). While the disk gas mass under the 50 K assumption is on the high end compared to similar Galactic sources, the computed $M_{gas}$ for 100 K is consistent, within errors, with the value of ~0.5 M$_\odot$ obtained from SED fitting[13]. However, the corresponding disk radius of about 100 AU obtained from the SED modeling is significantly smaller than the rotating structure observed here, supporting the likelihood of a radius of 6000 AU being a strict upper limit.

## Toomre stability analysis

We assess the stability of the structure consistent with Keplerian rotation (R ≲ 6000 AU, i.e., the disk) using the Toomre $Q$ instability parameter[36], which for Keplerian rotation is defined as,

$$Q = \frac{c_s \Omega}{\pi G \Sigma}$$

where $c_s$ is the sound speed, $\Omega = \sqrt{GM_{tot}/R^3}$ the angular velocity of the disk (with $M_{tot}$ the combined mass of the central star and the disk), and $\Sigma$ the surface density which is calculated as $\Sigma = M_{disk}/\pi R^2$. For temperatures of 100 K and 50 K (see above), a radius of 6000 AU, and adopting the derived disk and stellar masses, we obtain $Q = 3.5\pm1.8$ or $Q = 1.2\pm0.8$ (for 100 and 50 K, respectively). While both values are consistent with disk stability ($Q > 1$) we cannot exclude that the outermost disk radii are unstable, due to the large uncertainties stemming from both the adopted disk radius (which is a strict upper limit) and the assumed central star luminosity. However, given that the disk temperature is radius-dependent, as described above, the smaller radii become increasingly stable. If the radius of the disk is < 6000 AU, the disk temperature at the outer radii increases and the disk likely becomes stable at all radii. Detailed modeling of disks around massive stars at low metallicities will be necessary to quantitatively understand these types of systems.

**Competing financial interests**
The authors declare no competing financial interests.

**Data availability**
The authors declare that the data supporting the findings of this study are publicly available from the ALMA data archive under the project 2019.1.00756.S. All code generated for this study is available from the corresponding author upon request.

**Author Information**

The authors declare no competing financial interests. Correspondence and requests for materials should be addressed to A. F. M. ([anna.mcleod@durham.ac.uk](mailto:anna.mcleod@durham.ac.uk)).

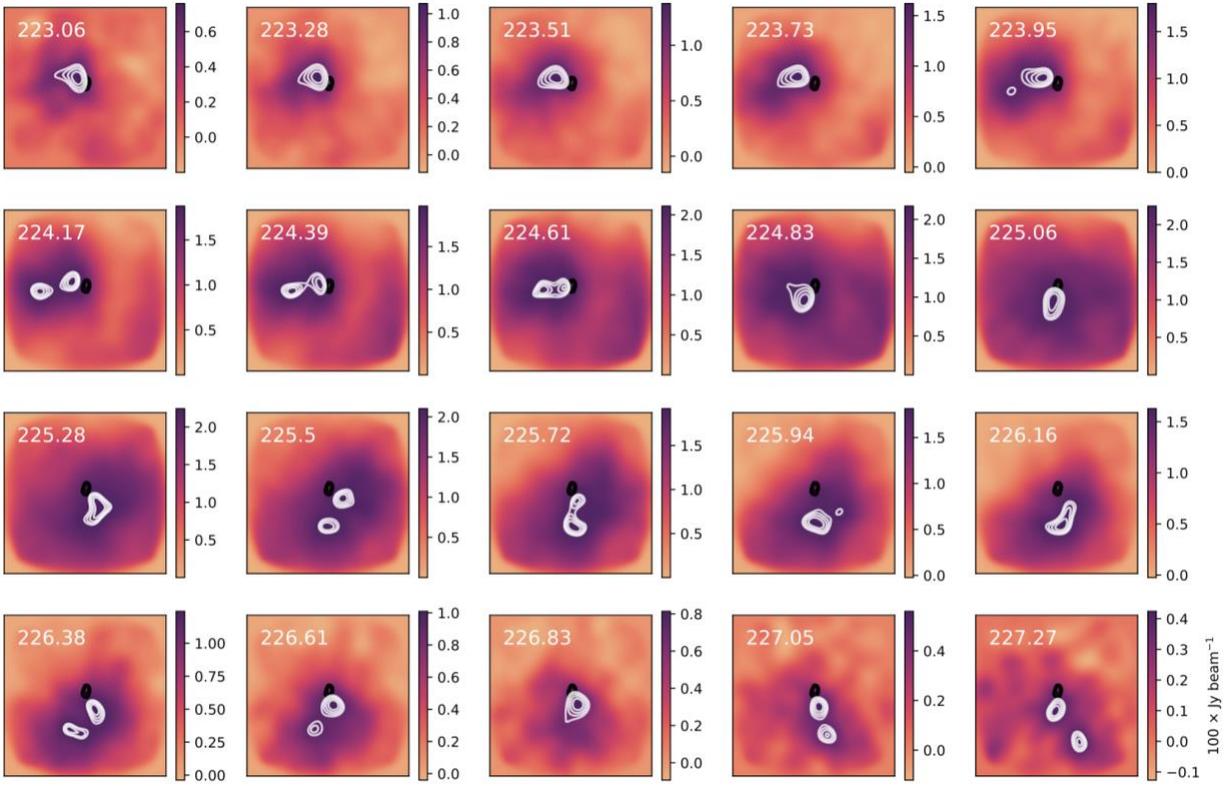

***Extended Data Figure 1. Channel maps of the*** ¹³***CO data cube***. *Velocities are indicated in each panel. Black contours show the continuum as in Figure 2, white contours trace the intensity of the colormaps (for simplicity, only ten contours of the top 3 percentile intensity values in each channel are shown).*

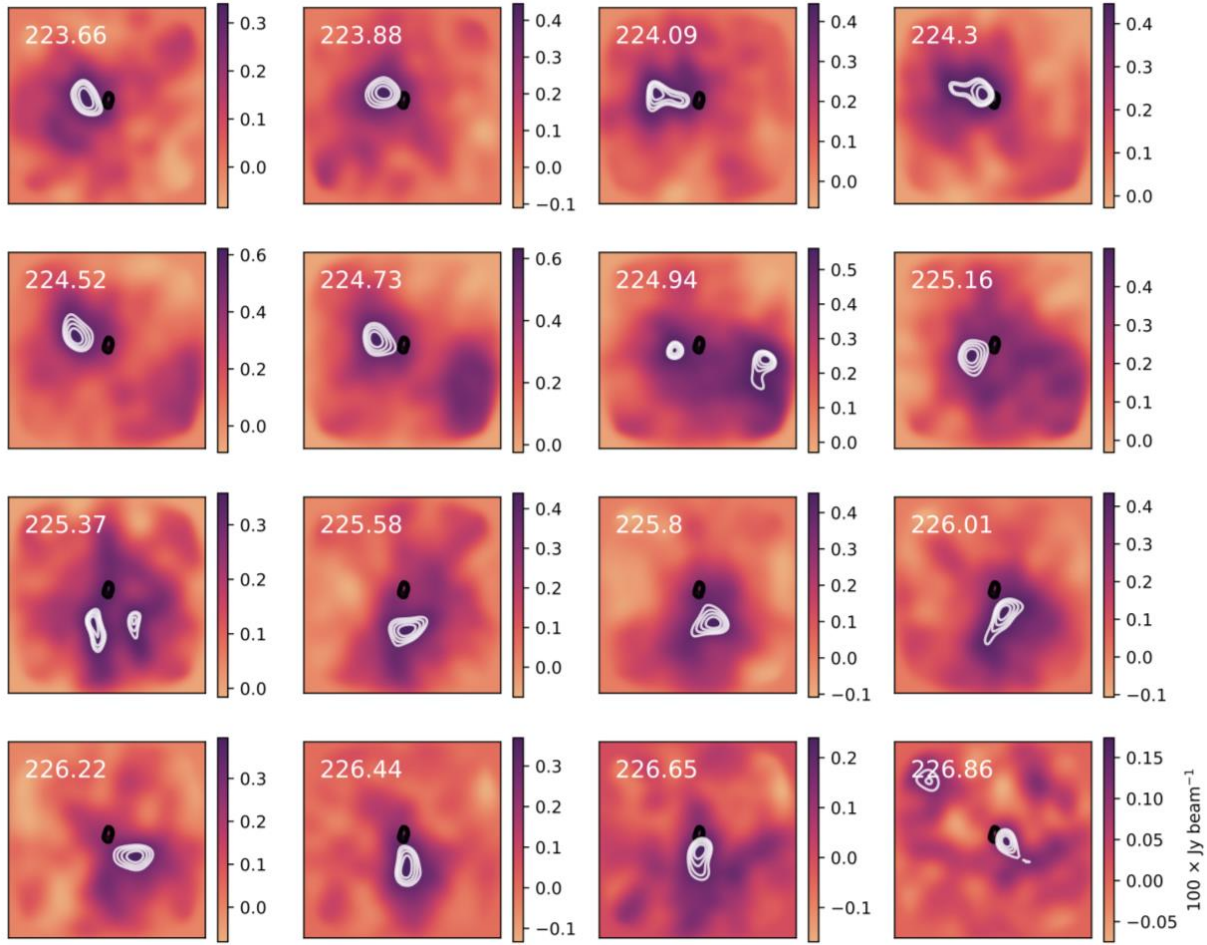

***Extended Data Figure 2. Channel maps of the CS data cube***. *Velocities are indicated in white in each panel. Black contours show the continuum as in Figure 2, white contours trace the intensity of the colormaps (for simplicity, only ten contours of the top 3 percentile intensity values in each channel are shown).*

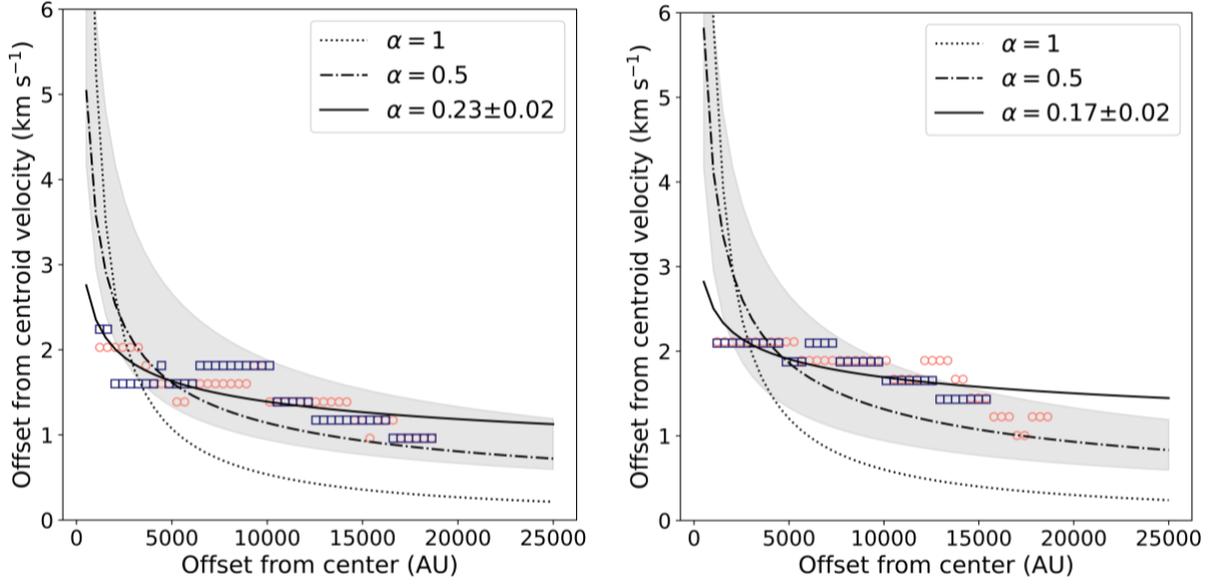

***Extended Data Figure 3. Best-fit power-laws to the gas kinematics***. *Best-fit power-laws to the redshifted (orange circles) and blueshifted (dark blue squares) radial velocity profiles of the outer envelope from Figure 2. Velocities are relative to the best-fit centroid velocity of the CS ($v_{LSR}$ = 225.3 km/s; left panel) and $^{13}CO$ ($v_{LSR}$ = 225.2 km/s; right panel) lines. The dotted and dot-dashed curves correspond to the best-fit power-laws with fixed index α = 1, and α = 0.5, respectively, while the solid curve corresponds to the best-fit power-law with α variable. The best-fit central source mass assuming Keplerian rotation (α = 0.5) is ~ 15 $M_\odot$ and ~ 19 $M_\odot$ for CS and $^{13}CO$, respectively. The shaded area indicates Keplerian rotation for the mass range 10 $M_\odot$ < M < 40 $M_\odot$.*

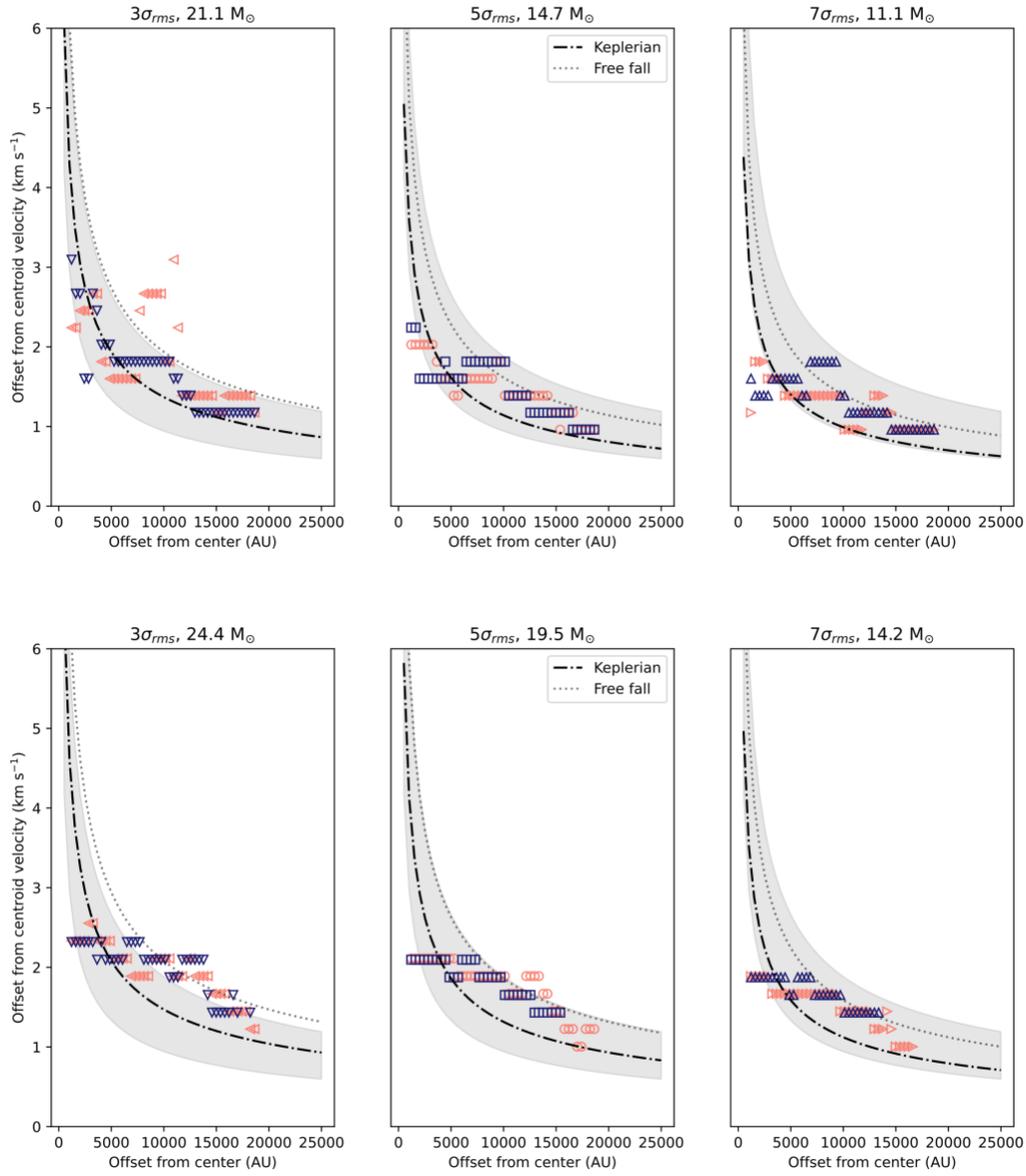

***Extended Data Figure 4. Rotation curves at different density thresholds.*** *Keplerian (dot-dashed) and free fall (dotted) curves for sources with central masses derived from the best fit $\alpha = 0.5$ power law to the redshifted (orange) and blueshifted (dark blue) radial velocity profiles of the outer emission envelope from Figure 3 (CS and $^{13}$CO in the upper and lower panels, respectively). The three panels correspond to different background noise thresholds used in determining the maximum rotation velocity[14] with the left showing a $3\sigma_{rms}$, the central panel the adopted $5\sigma_{rms}$, and the right a $7\sigma_{rms}$ threshold. Velocities are relative to the best-fit centroid velocity of the CS ($v_{LSR} = 225.3$ km/s) and $^{13}$CO ($v_{LSR} = 225.2$ km/s) lines. The shaded area indicates Keplerian rotation for the mass range $10 \, M_\odot < M < 40 \, M_\odot$.*